\documentclass[preprint,12pt]{elsarticle}

\usepackage{graphicx}

\usepackage{amssymb}

\biboptions{comma,square,numbers,sort&compress}

\journal{Physics Letters B}

\begin{document}

\begin{frontmatter}



\title{Measurement of the nuclear multiplicity ratio for $K^0_s$ hadronization at CLAS}





\newcommand{\ANL}{Argonne National Laboratory, Argonne, Illinois 60441}
\newcommand{\ANLindex}{1}
\newcommand{\ASU}{Arizona State University, Tempe, Arizona 85287-1504}
\newcommand{\ASUindex}{2}
\newcommand{\UCLA}{University of California at Los Angeles, Los Angeles, California  90095-1547}
\newcommand{\UCLAindex}{3}
\newcommand{\CSUDH}{California State University, Dominguez Hills, Carson, CA 90747}
\newcommand{\CSUDHindex}{4}
\newcommand{\CANISIUS}{Canisius College, Buffalo, NY}
\newcommand{\CANISIUSindex}{5}
\newcommand{\CMU}{Carnegie Mellon University, Pittsburgh, Pennsylvania 15213}
\newcommand{\CMUindex}{6}
\newcommand{\CUA}{Catholic University of America, Washington, D.C. 20064}
\newcommand{\CUAindex}{7}
\newcommand{\SACLAY}{CEA, Centre de Saclay, Irfu/Service de Physique Nucl\'eaire, 91191 Gif-sur-Yvette, France}
\newcommand{\SACLAYindex}{8}
\newcommand{\CNU}{Christopher Newport University, Newport News, Virginia 23606}
\newcommand{\CNUindex}{9}
\newcommand{\UCONN}{University of Connecticut, Storrs, Connecticut 06269}
\newcommand{\UCONNindex}{10}
\newcommand{\EDINBURGH}{Edinburgh University, Edinburgh EH9 3JZ, United Kingdom}
\newcommand{\EDINBURGHindex}{11}
\newcommand{\FU}{Fairfield University, Fairfield CT 06824}
\newcommand{\FUindex}{12}
\newcommand{\FIU}{Florida International University, Miami, Florida 33199}
\newcommand{\FIUindex}{13}
\newcommand{\FSU}{Florida State University, Tallahassee, Florida 32306}
\newcommand{\FSUindex}{14}
\newcommand{\Genova}{Universit$\grave{a}$ di Genova, 16146 Genova, Italy}
\newcommand{\Genovaindex}{15}
\newcommand{\GWUI}{The George Washington University, Washington, DC 20052}
\newcommand{\GWUIindex}{16}
\newcommand{\ISU}{Idaho State University, Pocatello, Idaho 83209}
\newcommand{\ISUindex}{17}
\newcommand{\INFNFE}{INFN, Sezione di Ferrara, 44100 Ferrara, Italy}
\newcommand{\INFNFEindex}{18}
\newcommand{\INFNFR}{INFN, Laboratori Nazionali di Frascati, 00044 Frascati, Italy}
\newcommand{\INFNFRindex}{19}
\newcommand{\INFNGE}{INFN, Sezione di Genova, 16146 Genova, Italy}
\newcommand{\INFNGEindex}{20}
\newcommand{\INFNRO}{INFN, Sezione di Roma Tor Vergata, 00133 Rome, Italy}
\newcommand{\INFNROindex}{21}
\newcommand{\ORSAY}{Institut de Physique Nucl\'eaire ORSAY, Orsay, France}
\newcommand{\ORSAYindex}{22}
\newcommand{\ITEP}{Institute of Theoretical and Experimental Physics, Moscow, 117259, Russia}
\newcommand{\ITEPindex}{23}
\newcommand{\JMU}{James Madison University, Harrisonburg, Virginia 22807}
\newcommand{\JMUindex}{24}
\newcommand{\KNU}{Kyungpook National University, Daegu 702-701, Republic of Korea}
\newcommand{\KNUindex}{25}
\newcommand{\LPSC}{LPSC, Universite Joseph Fourier, CNRS/IN2P3, INPG, Grenoble, France}
\newcommand{\LPSCindex}{26}
\newcommand{\UNH}{University of New Hampshire, Durham, New Hampshire 03824-3568}
\newcommand{\UNHindex}{27}
\newcommand{\NSU}{Norfolk State University, Norfolk, Virginia 23504}
\newcommand{\NSUindex}{28}
\newcommand{\OHIOU}{Ohio University, Athens, Ohio  45701}
\newcommand{\OHIOUindex}{29}
\newcommand{\ODU}{Old Dominion University, Norfolk, Virginia 23529}
\newcommand{\ODUindex}{30}
\newcommand{\RPI}{Rensselaer Polytechnic Institute, Troy, New York 12180-3590}
\newcommand{\RPIindex}{31}
\newcommand{\URICH}{University of Richmond, Richmond, Virginia 23173}
\newcommand{\URICHindex}{32}
\newcommand{\ROMAII}{Universita' di Roma Tor Vergata, 00133 Rome Italy}
\newcommand{\ROMAIIindex}{33}
\newcommand{\MSU}{Skobeltsyn Nuclear Physics Institute, Skobeltsyn Nuclear Physics Institute, 119899 Moscow, Russia}
\newcommand{\MSUindex}{34}
\newcommand{\SCAROLINA}{University of South Carolina, Columbia, South Carolina 29208}
\newcommand{\SCAROLINAindex}{35}
\newcommand{\JLAB}{Thomas Jefferson National Accelerator Facility, Newport News, Virginia 23606}
\newcommand{\JLABindex}{36}
\newcommand{\UNIONC}{Union College, Schenectady, NY 12308}
\newcommand{\UNIONCindex}{37}
\newcommand{\UTFSM}{Universidad T\'{e}cnica Federico Santa Mar\'{i}a, Casilla 110-V Valpara\'{i}so, Chile}
\newcommand{\UTFSMindex}{38}
\newcommand{\GLASGOW}{University of Glasgow, Glasgow G12 8QQ, United Kingdom}
\newcommand{\GLASGOWindex}{39}
\newcommand{\VIRGINIA}{University of Virginia, Charlottesville, Virginia 22901}
\newcommand{\VIRGINIAindex}{40}
\newcommand{\WM}{College of William and Mary, Williamsburg, Virginia 23187-8795}
\newcommand{\WMindex}{41}
\newcommand{\YEREVAN}{Yerevan Physics Institute, 375036 Yerevan, Armenia}
\newcommand{\YEREVANindex}{42}

\newcommand{\NOWMSU}{Skobeltsyn Nuclear Physics Institute, Skobeltsyn Nuclear Physics Institute, 119899 Moscow, Russia}
\newcommand{\NOWINFNGE}{INFN, Sezione di Genova, 16146 Genova, Italy}
\newcommand{\NOWNONE}{unknown, No Adress Available}

\author[toOHIOU]{A.~Daniel}
\author[toOHIOU]{K.~Hicks}
\author[toUTFSM,toJLAB]{W.K.~Brooks}
\author[toUTFSM,toYEREVAN]{H.~Hakobyan}
\author[toODU]{K.P. ~Adhikari}
\author[toODU]{D.~Adikaram}
\author[toINFNFR]{M.~Aghasyan}
\author[toODU]{M.~Amarian}
\author[toINFNGE]{M.~Anghinolfi}
\author[toJLAB]{H.~Avakian}
% ...



% ...
\address[toINFNFR]{\INFNFR} 
 \address[toINFNGE]{\INFNGE} 
% ...
\address[toOHIOU]{\OHIOU} 
 \address[toODU]{\ODU} 
% ...
\address[toJLAB]{\JLAB} 
% ...
\address[toUTFSM]{\UTFSM} 
% ...
\address[toYEREVAN]{\YEREVAN}

 \fntext[toNOWMSU]{Current address: Skobeltsyn Nuclear Physics Institute, 119899 Moscow, Russia }
 \fntext[toNOWINFNGE]{Current address: INFN, Sezione di Genova, 16146 Genova, Italy }

\begin{abstract}
The influence of cold nuclear matter on lepto-production  of hadrons 
in semi-inclusive deep inelastic scattering is measured using the CLAS 
detector in Hall B at Jefferson Lab and a 5.014 GeV electron beam. 
We report the $K_s^0$ multiplicity ratios for targets of C, Fe, and Pb relative 
to deuterium as a function of the fractional virtual
 photon energy $z$ transferred to the $K_s^0$ and the transverse momentum squared $p_{T}^2$ of
    the $K_s^0$. We find that the multiplicity ratios for $K^0_s$ are reduced in the 
nuclear medium at high $z$ and low $p_{T}^2$, with a trend for the 
$K^0_s$ transverse momentum to be broadened in the nucleus for large $p_{T}^2$.
\end{abstract}


\end{frontmatter}
\newcommand{\rha}{$R^h_A$ }  
\newcommand{\dptsq}{$\Delta{p_{T}^{2}}$ }
\newcommand{\ks}{$K^0_s$ }
\newcommand{\ptsq}{$p_{T}^{2}$ }


Hadronization is the process through which partons, created in an elementary reaction, turn into hadrons. An example for this process is lepto-production of hadrons in semi-inclusive deep-inelastic electron-nucleon scattering, where a parton is struck by a virtual photon; as the parton propagates away from the interaction point it picks up partner quarks (antiquarks) to form a hadron. The process of hadronization is changed by the nuclear medium. Nuclear modification of hadron production in deep-inelastic scattering (DIS) was first observed at 
SLAC~\cite{SLAC_had} followed by EMC~\cite{EMC_had}, E665~\cite{E665_had} 
and more recently at  HERMES~\cite{hermes2001,hermes2003,hermes2007}. The goal of these investigations is to study the mechanism of quark deconfinement followed by propagation and then reconfinement into a 
hadron, and in particular how the presence of cold nuclear matter 
affects the hadronization process.
Although some rescattering of the quark/hadron in nuclear matter is 
expected, calculations \cite{gallmeister,kopeliovich} for the 
HERMES data suggest that the effects are more than just simple 
rescattering--the hadronization mechanism itself is changed 
due to the propagation through nuclear matter. In addition to 
understanding the fundamental features of hadronization, a detailed 
understanding of the nuclear modifications is important for the 
interpretation of ultra-relativistic heavy ion collisions,  
as well as the study of neutrino oscillations using 
nuclear targets in the GeV energy range. A recent review on 
the subject of hadronization and parton propagation can be found in
reference~\cite{Accardi:2009review}.

The observables reported in the present paper are the multiplicity ratios as a function 
of $z=E_h/\nu$, the fraction of the virtual-photon energy carried by 
the hadron in the target rest frame, and its transverse momentum squared, $p_T^2$. The transverse momentum is defined as the perpendicular component of the hadron's momentum measured with respect to the 
direction of momentum transfer in  electron scattering.
The multiplicity ratio \rha is defined as the ratio of the number of 
hadrons detected, normalized to the number of electrons measured in 
the DIS kinematics for a nuclear target with mass number A, divided by the same quantities 
for a deuterium target(D);
\begin{equation}
R^{h}_{A}= \frac
{\left[N_{h}\left( z,p_{T}^{2},Q^{2},\nu\right)
/ N_{e^{-}}^{DIS}\left( \nu,Q^{2}\right)\right]_{A}} 
{\left[N_{h}\left( z,p_{T}^{2},Q^{2},\nu \right)
/ N_{e^{-}}^{DIS}\left( \nu,Q^{2}\right)\right]_{D} },
\label{eq:rhm}
\end{equation}
where $N_{h}$ is the number of hadrons  and $N_{e^{-}}^{DIS}$ is the 
number of scattered electrons detected in the specified kinematic bins. Here, $Q^2$ is the negative of the four-momentum of the virtual photon squared and $\nu$ is the energy of the virtual photon. For most of the explored  kinematic range \rha shows a reduction from unity, although at low $z$ and high $p_{T}^2$ it shows an enhancement. 
Hadron \textit{formation lengths}, the characteristic distances over which 
hadrons form, can be extracted from $R^h_A$ \cite{bialas}. 

 In certain theoretical models, the transverse momentum is broadened 
for hadronization from nuclear targets as compared with the 
distribution measured from deuterium, and the amount of broadening 
is sensitive to gluonic radiation by the quark before it evolves into 
a hadron \cite{bdmps, Accardi:2009review, hermes_ptbroad2010}. 
By varying the nuclear radius, measurements of the 
broadening as a function of the kinematic variables allow one 
to infer the length scale (known as the \textit{production length}) 
over which the quark is deconfined. 

Since hadronization is a nonperturbative QCD process that as yet cannot 
be calculated from first principles, one must rely on  models 
to interpret the observed nuclear dependence of hadronization. 
In particular, several hadron species are required to analyze the 
flavor dependence of the observables, to unravel the reaction 
mechanisms involved and to give insight into the fundamental principles 
governing hadron formation. Of special interest is the hadronization into 
strange mesons as compared with non-strange mesons.  In the HERMES data 
\cite{hermes2007}, the $z$ dependence of $R^h_A$ in the medium for $\pi^+$,
$\pi^-$ and $\pi^0$ is nearly identical for a given nucleus.  However, $R^h_A$ for $K^{+}$ is  systematically larger than for $K^{-}$. The difference for the charged kaons can be explained due to the larger nuclear absorption for $K^{-}$ compared to  $K^{+}$. Also, note  that  pions have only non-strange valence quarks, $K^+$  has a non-strange valence quark while  $K^-$ has a non-strange antiquark and strange quark content.  Thus, the production of a  negative kaon requires further complicated string breaking mechanisms  and fewer leading hadrons are expected to be $K^{-}$ as compared to  $K^{+}$.


 While the HERMES measurements for the hadronization effects in $\pi^0$ production looked very similar to $\pi^\pm$, they did not report any $K^0$ hadronization results.  
Strangeness production can give a unique insight into the hadronization 
process, since the strange quark is likely to come directly from the 
``string-breaking'' process \cite{hermes2007}.  Different hadron species, such as the 
$K^0_s$, are relevant to the study of the flavor dependence of 
hadronization and the mass dependence of quark energy loss. 
In this letter, we report for the first time on $K_s^0$ hadronization measured using the CLAS detector at Jefferson Lab. Our results 
show, like the HERMES results, that the hadronization mechanism is 
affected by the presence of the nuclear medium. Of course, the $K_s^0$ is a 
mixture of $K^0$ and $\bar{K}^0$, but just as $K^+$ hadronization 
dominates over $K^-$ (presumably due to the non-strange valence quark in the 
$K^+$) we expect that most of the $K_s^0$ signal is dominated by 
$K^0$ hadronization, with a smaller contribution from $\bar{K}^0$.  

The experiment was carried out using the CLAS detector~\cite{NIM:CLAS} 
at the Thomas Jefferson National Accelerator Facility. 
The data presented here are from the E02-104~\cite{qp_6gevprop} 
experiment. An electron beam of 5.014 GeV 
was incident on targets of liquid deuterium, solid carbon, iron and 
lead.  The targets were arranged in a mechanical structure that 
allowed one solid nuclear target to be in the beam at the same time 
as (and slightly downstream of) the liquid deuterium target~\cite{eg2tar_nim}. The length of liquid target was about 2 cm  and the separation distance between solid and liquid targets was approximately 4 cm.  This 
allowed for reduced systematic uncertainties in the ratios of hadrons 
measured by the CLAS detector. The scattered electron and the produced 
hadrons were detected in coincidence by the CLAS detector. 
Details of the experiment are described in Ref.~\cite{Hyke:thesis}. 

\begin{figure}[htbp]
\centering
\begin{tabular}{cc}
\includegraphics[height=7.0cm,width=6.5cm]{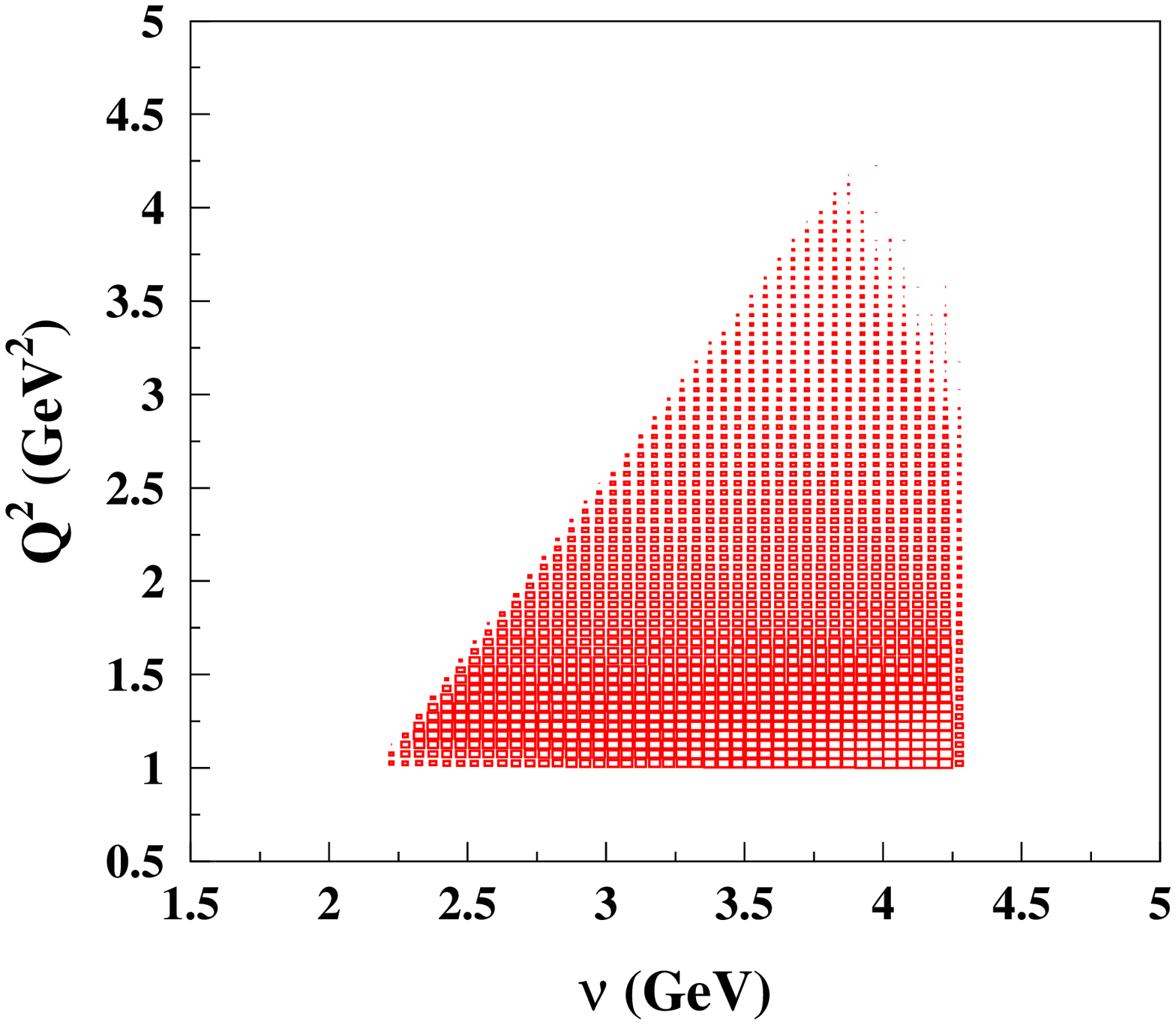}&
\includegraphics[height=7.0cm,width=6.5cm]{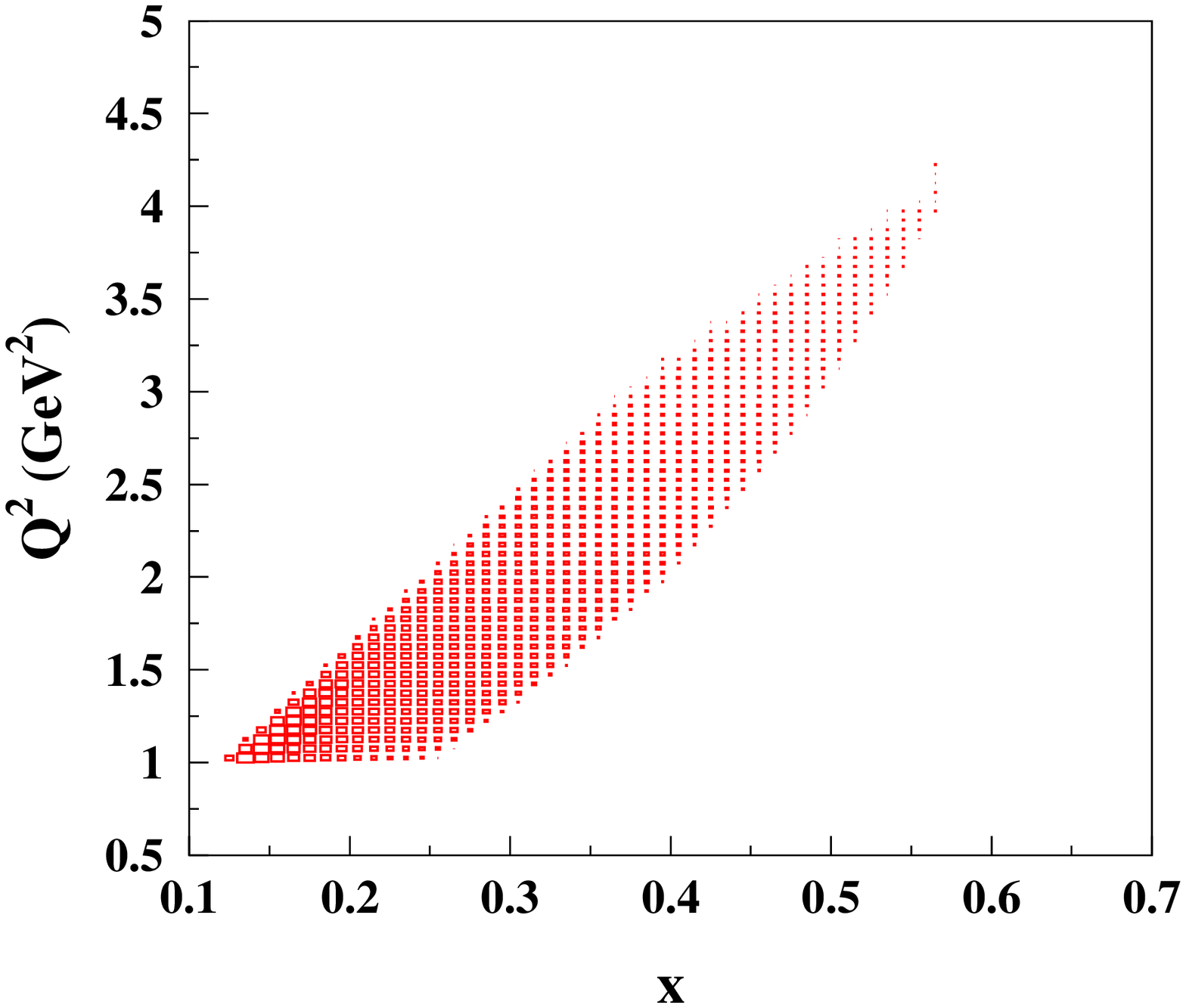}
\end{tabular}
\linespread{0.5}
\caption [] {Distribution of $Q^2$ versus $\nu$ (left) and $x$ (right), for DIS events 
from a 5.014 GeV electron beam on a liquid deuterium target, after the kinematic 
cuts described in the text.
\label{fig:q2dep}}
\end{figure}

The DIS kinematics covered by the present experiment are shown in 
Fig.~\ref{fig:q2dep}, where  $Q^2$ is plotted versus  $\nu$ and also versus Bjorken $x=Q^2/(2M\nu)$, where $M$ the nucleon mass. 
Constraints have been applied, limiting the kinematic variables to 
 $Q^2 > 1.0$ GeV$^2$,  $W^2 > 4.0$ GeV$^2$ and $y< 0.85$, 
where $W^2$ is the squared invariant mass of the photon-nucleon system, $y=\nu/E$ the energy fraction of the virtual photon in the target 
rest frame, and $E$ is the beam energy.  The constraint on $y$ is applied to reduce the magnitude of radiative corrections. The range in $x$ depends on $Q^2$ and extends from $x=0.1$ at the lowest 
$Q^2$ up to $x=0.55$ at the highest $Q^2$.  Though the sea quarks may have non-negligible contribution at the smallest $x$ values probed in this experiment, for $x$ greater than about 0.3 the virtual photon is likely to strike a valence quark. Note that the kinematics of the CLAS results are  
different from those of HERMES results, which had higher $\nu$ and 
hence lower $x$, extending down to 0.023. 
\begin{figure}[htbp]
\centering
\begin{tabular}{cc}
\includegraphics[height=5.5cm,width=5.5cm]{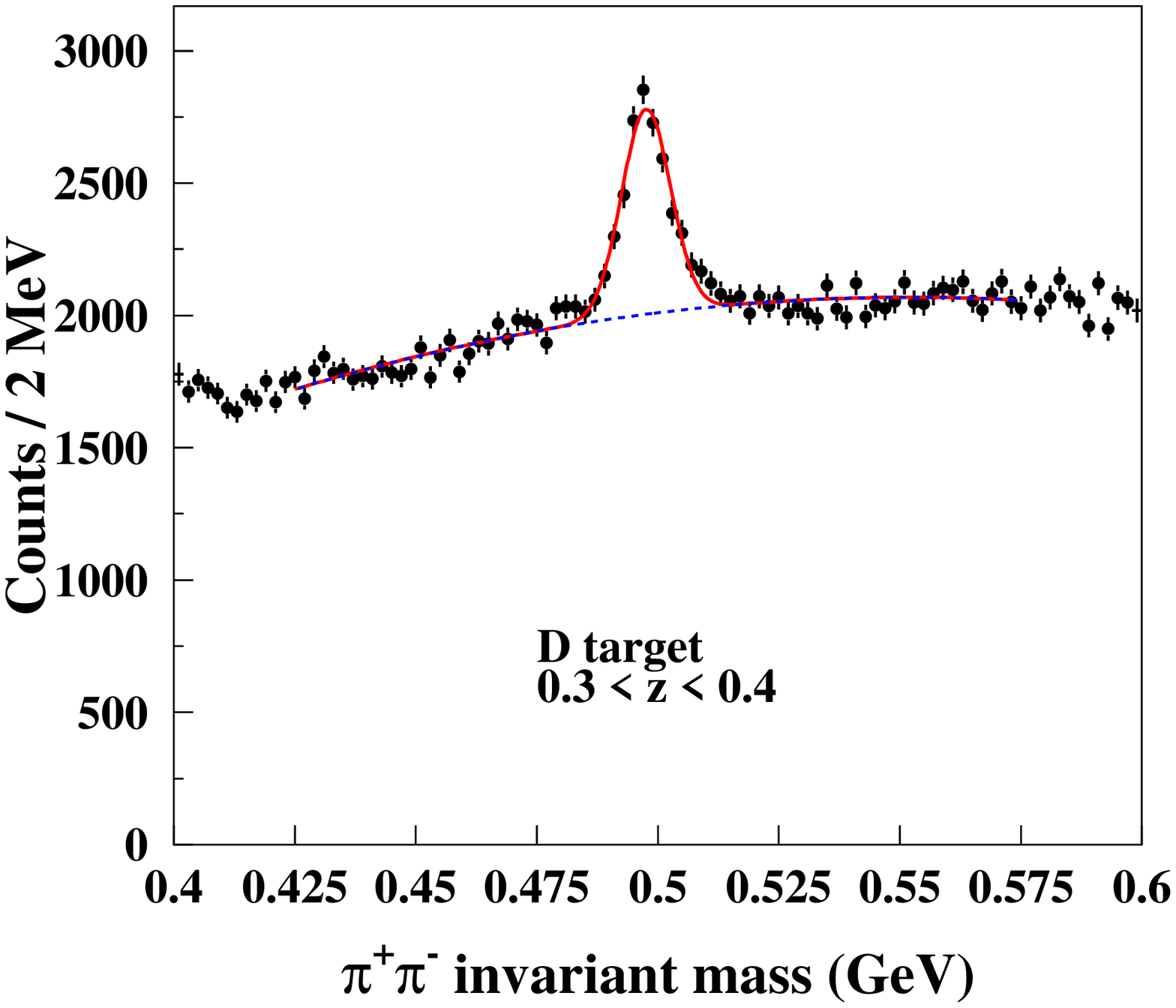}&
\includegraphics[height=5.5cm,width=5.5cm]{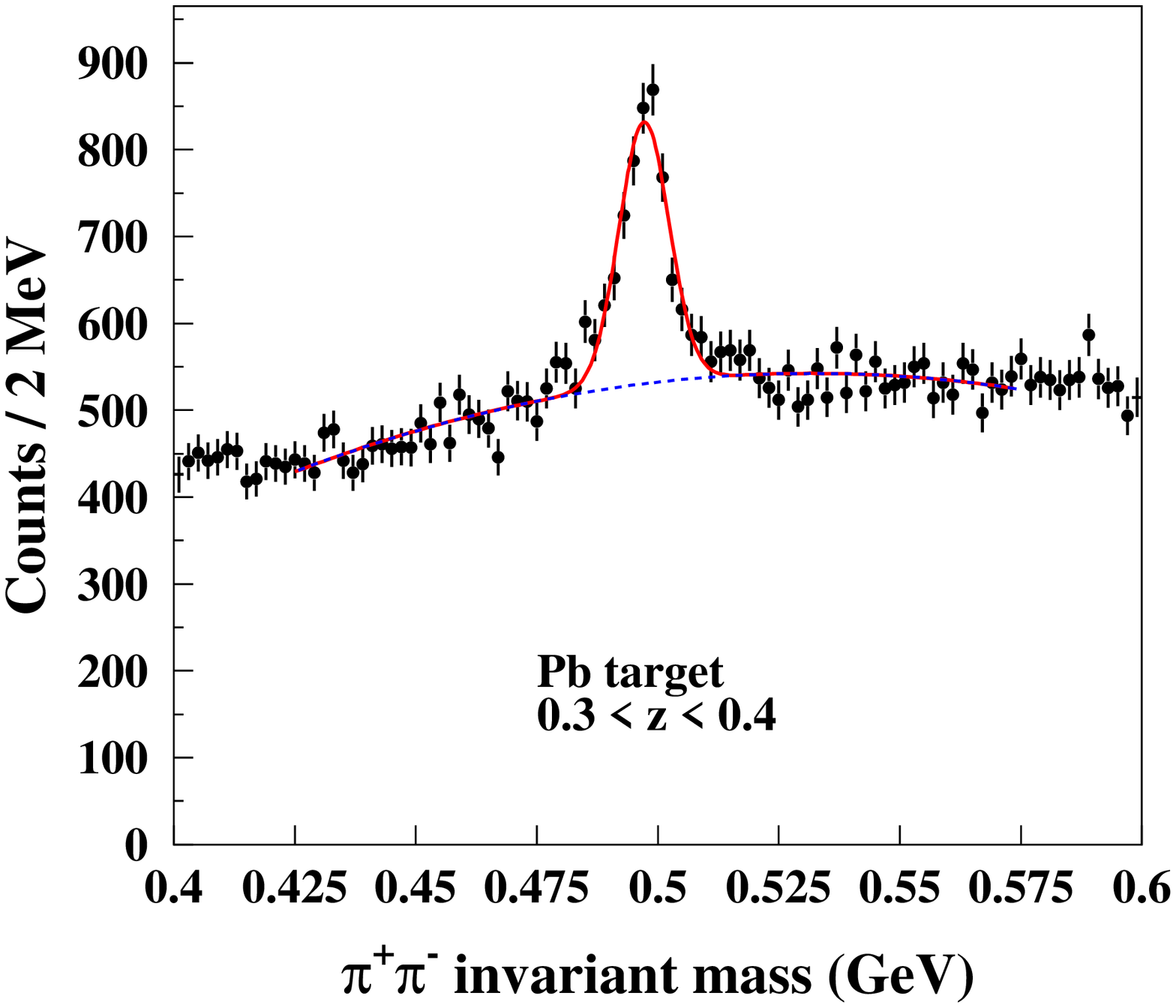}
\\
\includegraphics[height=5.5cm,width=5.5cm]{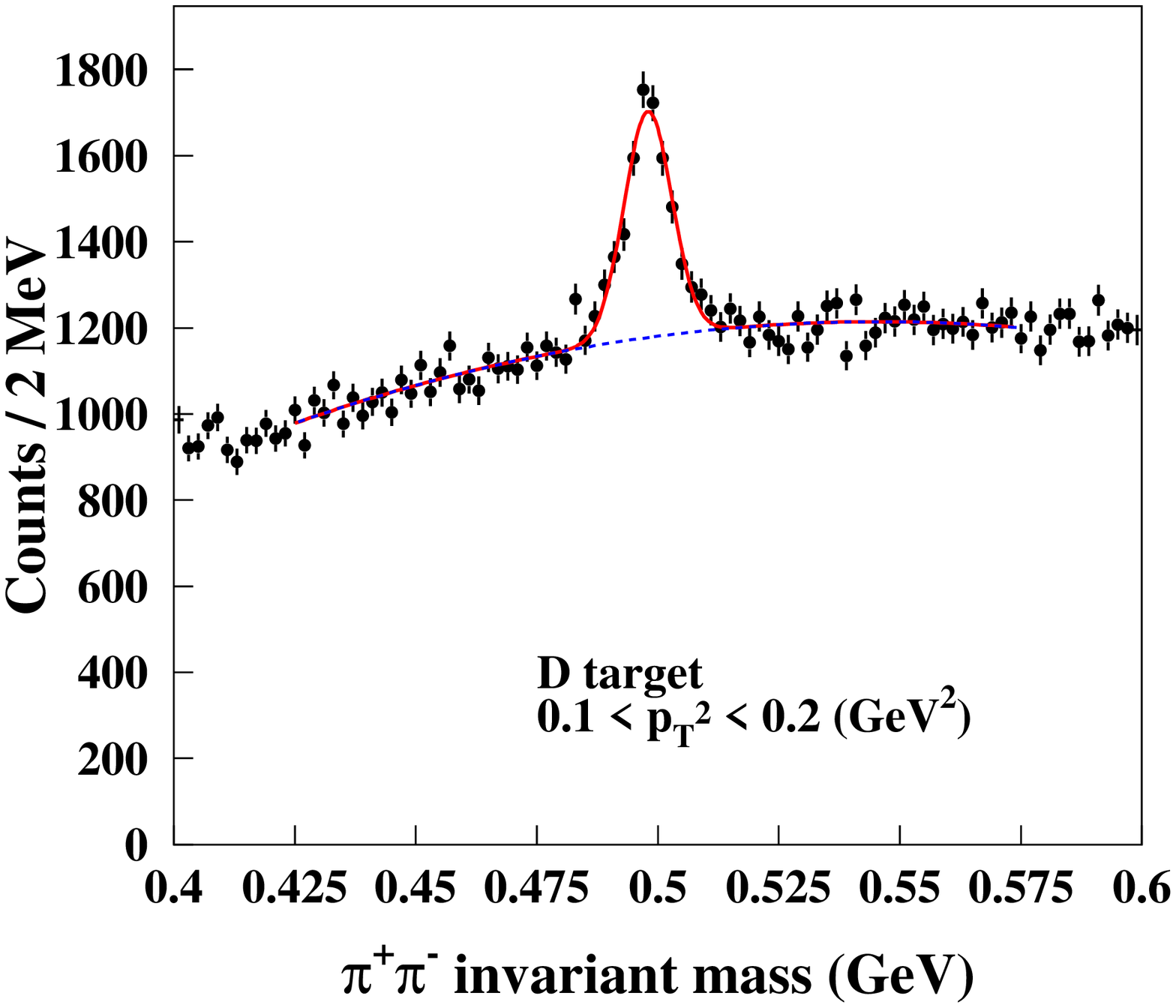}&
\includegraphics[height=5.5cm,width=5.5cm]{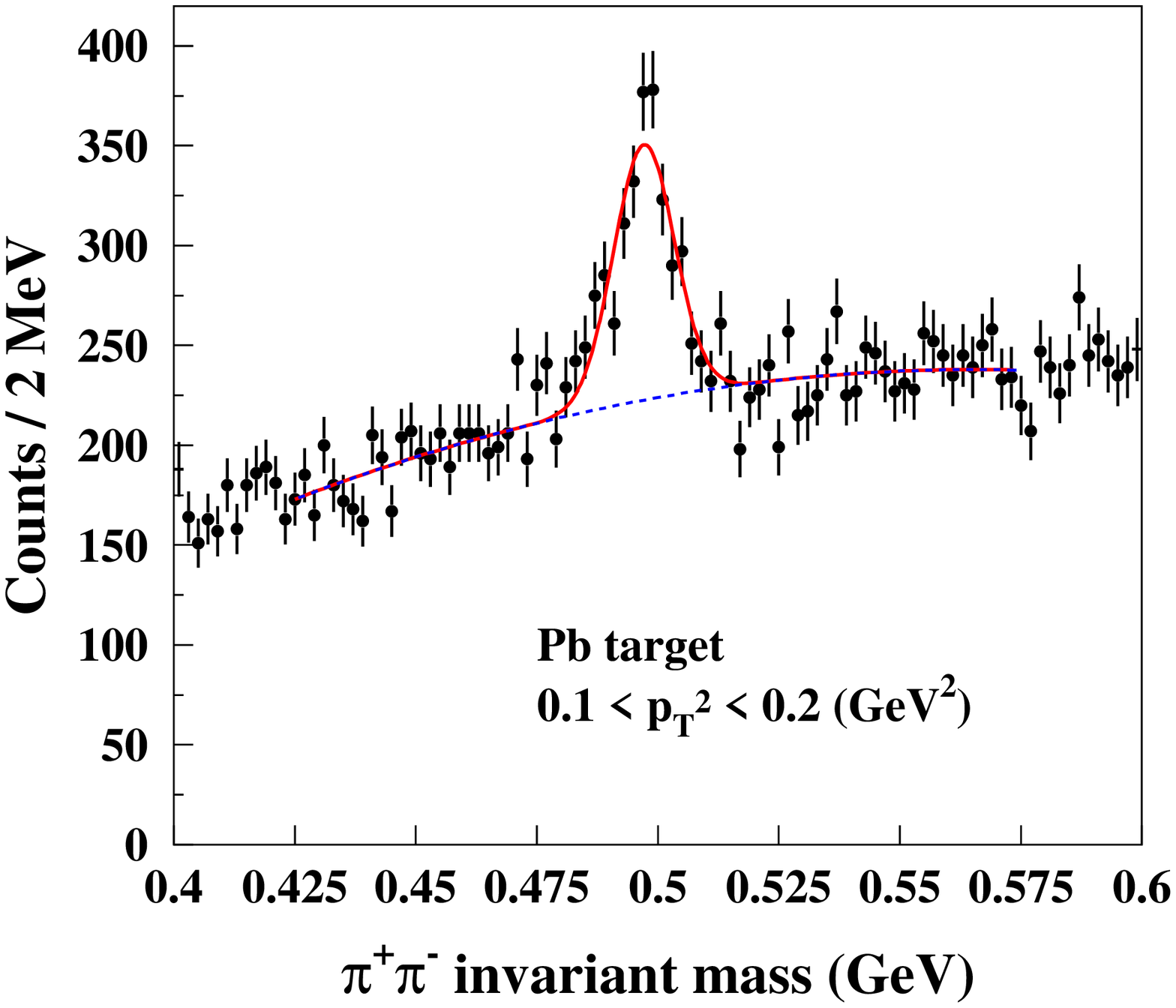}
\end{tabular}
\linespread{0.5}
\caption [] {Invariant mass of two oppositely charged pions, showing a 
peak at the $K^0$ mass, for the liquid deuterium target (left) and the 
solid Pb target (right) for selected kinematic bins: $0.3<z<0.4$ (top) 
and $0.1<p_T^2<0.2$ GeV$^2$ (bottom). The $x$-bin size is 2 MeV.
\label{fig:invmass}}
\end{figure}

The $K^0_s$'s were found from fits to the invariant mass distribution of 
$\pi^+\pi^-$ pairs produced. After identifying the scattered electron, $K^0_s$ invariant mass spectra $M(\pi^+,\pi^-)$ is formed from the combination of all charged pion pairs, detected in CLAS. For a 
given bin in $z$, and integrated over all electron kinematics and all other hadron variables, 
a clear peak in $M(\pi^+,\pi^-)$ is seen at the mass of the $K^0_s$ as 
shown in the top plots of Fig.~\ref{fig:invmass} for both the liquid 
deuterium (left) and solid Pb (right) targets. 
Similarly, plots showing the $K^0_s$ peaks for a bin in $p_T^2$ after integrated over all electron kinematics and all other hadron variables are shown in the bottom plots of Fig.~\ref{fig:invmass}. 
Although substantial, the  background from uncorrelated $\pi^+\pi^-$ 
pairs is smooth and is easily fit by a polynomial.  
The $K^0_s$ peak was fit using an unconstrained Gaussian. 
The statistical uncertainties are determined directly from the fit parameters.

\begin{figure}[htb]
\centering
\begin{tabular}{cc}
\includegraphics[height=8.0cm,width=7.2cm]{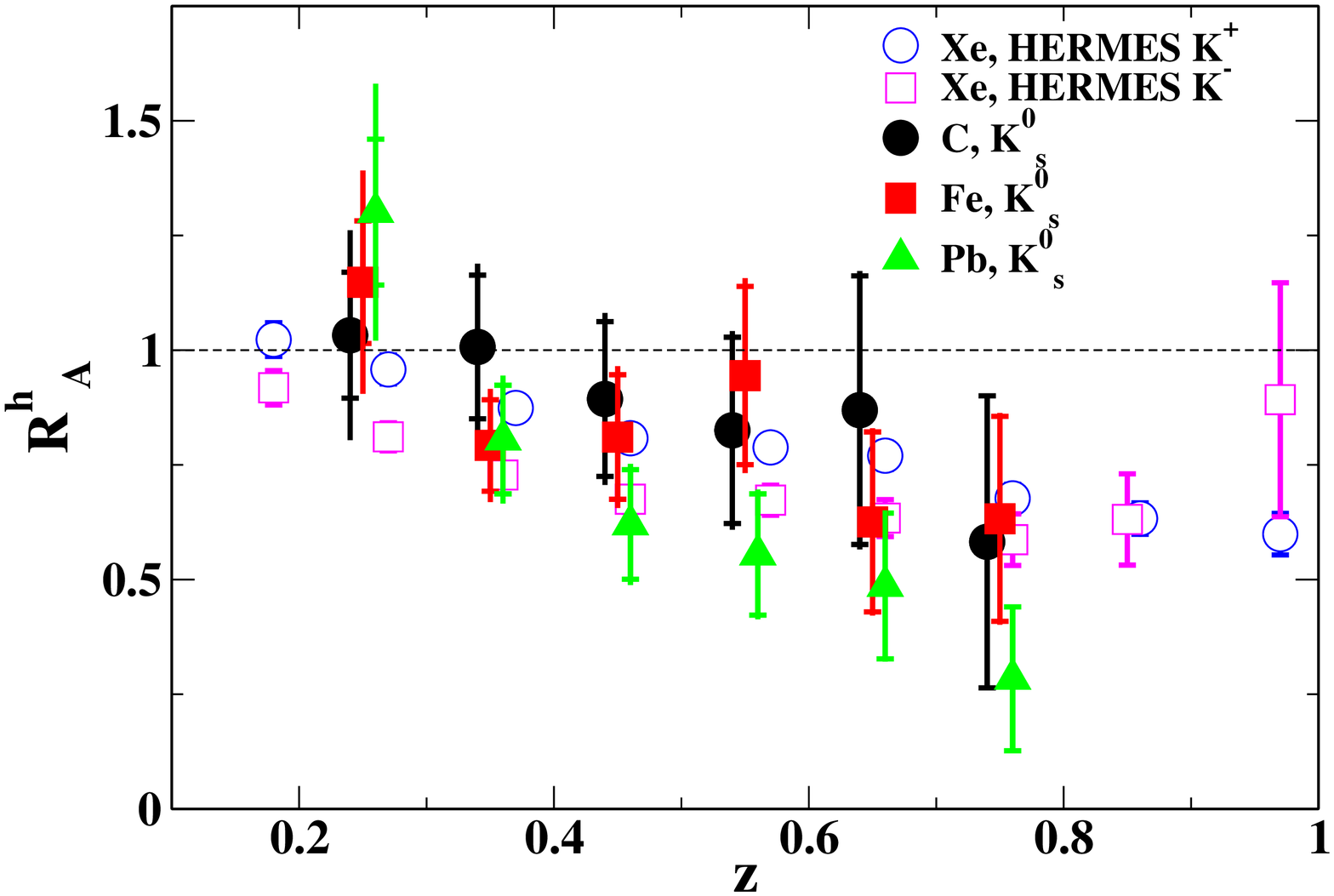}&
\includegraphics[height=8.0cm,width=7.2cm]{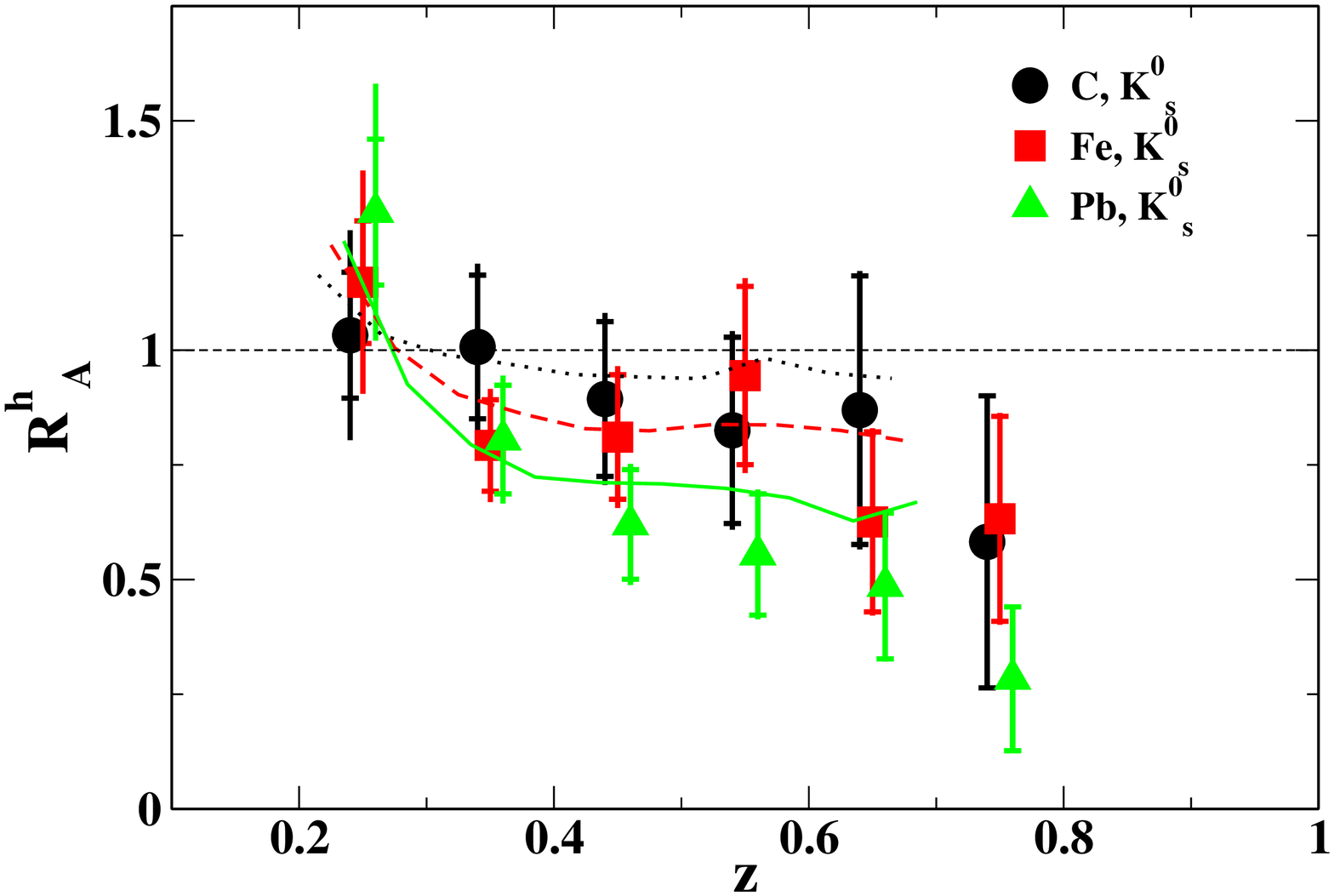}
\end{tabular}
\linespread{0.5}
\caption [] {Left panel shows the CLAS data from the EG2 experiment for the hadronic multiplicity ratio for \ks versus  $z$ along with HERMES results \cite{hermes2007} for charged kaons for the Xe nuclei. Right  panel shows the CLAS results  along with the results derived from calculations of Ref. \protect\cite{gallmeister} for Pb (solid line), Fe (dashed line) and C (dotted line). Data points are shown with different symbols as given  in the legend. The inner error bars represent the statistical uncertainty, while the outer ones show the total uncertainty. For clarity, the HERMES points are shifted by +0.02 units in $z$ and the C and Pb  points are shifted by -0.01 and +0.01 units in $z$, respectively.
\label{fig:k0mult_z}}
\end{figure}

\begin{figure}[htb]
\begin{center}
\includegraphics[height=9.0cm]{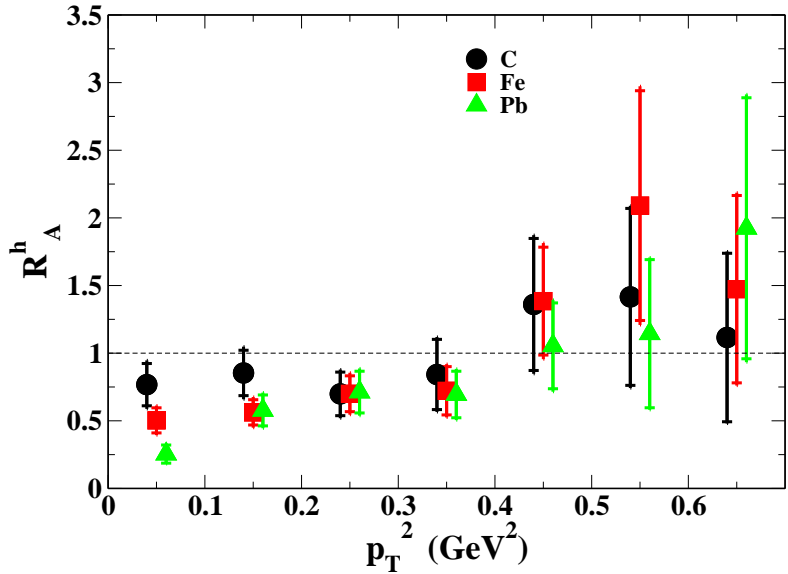}
\caption [] {CLAS data for the hadronic multiplicity ratio for \ks as a 
function of $p_T^2$ for $0.3<z<0.8$. Error bars are as in Fig.~\ref{fig:k0mult_z}. For clarity, the C and Pb  points are shifted by -0.01 and +0.01 units in $p_T^2$, respectively.} 
\label{fig:k0mult_pt2}
\end{center}
\end{figure}

Results for $R^h_A$ as a function of $z$ and of $p_T^2$ 
for $K^0_s$ hadronization are shown in 
Figs.~\ref{fig:k0mult_z} and \ref{fig:k0mult_pt2}, respectively, 
for the carbon, iron and lead targets. The kinematic 
dependence on a given variable is extracted  while integrating over 
all other variables within the acceptance of the experiment. Results for $R^h_A$ vs. $z$ for the charged kaons from the HERMES experiment are also shown in left panel of Fig.~\ref{fig:k0mult_z}. For comparison, we have plotted only the results for the xenon target from the HERMES experiment; results for several nuclei and different hadrons are available in Ref.~\cite{hermes2007}. For the \ptsq dependence studies,  the data presented are limited to  
$0.3<z<0.8$ to reduce the possible contamination from  target 
fragmentation and  from exclusive channels. The ratios of $K^0_s$ have 
been normalized by the number of DIS electrons for each target and corrected for geometrical detector acceptance for each target as calculated by Monte Carlo simulations using a PYTHIA event generator~\cite{pythia, Hyke:thesis}. The acceptance correction was applied on a bin-by-bin basis. No corrections have been applied to the 
data for  QED radiative effects, which were too small to measure 
with the present statistics. However, our  studies showed that 
these corrections are less than 5\% for the multiplicity ratios and this 
is included in the systematic  uncertainty. 

Several sources contribute to the overall systematic uncertainty in the multiplicity ratios.  Details of the systematic uncertainty studies are described in Ref.~\cite{CLAS:note}. The systematic uncertainty in the acceptance correction was found to be less than 5\% for the $z$ dependence (except for the lowest $z$ bin where this uncertainty is estimated to be 16\%)  and less than 8\% for the $p_{T}^{2}$ dependence. The systematic uncertainty associated with particle identification for the multiplicity ratios was found to be less than 3\%. To account for the possible systematic uncertainty in the fitting procedures we have assigned a 4\% uncertainty in  $R^h_A$ vs. $z$ results and a 5\%  uncertainty in $R^h_A$ vs. $p_T^2$. The  total uncertainty is obtained by adding systematic and statistical uncertainties in quadrature, and are shown by the longer error bars in Figs.~\ref{fig:k0mult_z} and \ref{fig:k0mult_pt2}. 

Fig.~\ref{fig:k0mult_z} shows that, for $z>0.3$, fewer $K^0_s$ are 
seen from nuclear targets (normalized by the number of DIS events) 
than from deuterium while at $z<0.3$ the ratio shows a trend to be bigger than unity. At lower $z$, effects such as nuclear rescattering, 
including $(\pi,K)$ reactions, and target fragmentation are likely to 
contribute to the number of $K^0_s$ measured. The same effect at low 
$z$ for $K^+$ hadronization has already been noted by HERMES \cite{hermes2003}.
The trend of the CLAS $K^0_s$ hadronization data is similar to that of 
the HERMES results, where more attenuation is seen as $z$ increases, with 
larger attenuation factors for the heaviest target.  
The present uncertainties are large due to the limited statistics, however the data agree reasonably well with predictions \cite{gallmeister} as shown in right panel of Fig.~\ref{fig:k0mult_z}. 

The theoretical calculations shown in Fig.~\ref{fig:k0mult_z} and presented in Ref.~\cite{gallmeister} are carried out in the framework of a probabilistic coupled-channel transport model based on the Boltzmann-Uehling-Uhlenbeck equation, which allows for detailed treatment of the final-state interactions beyond simple absorption mechanisms. It starts with the Lund string fragmentation picture as embodied by PYTHIA ~\cite{pythia} and extracts two points for each event: (i) the location in space where the struck quark breaks the string, and (ii) the location where the quark joins with an antiquark to form a meson. This is done event-by-event, in the context of a Monte Carlo calculation.  Medium effects are included by assuming a phenomenological transition from a pre-hadronic scattering cross section to a  full cross section constrained by hadronic beam data.  A function linear in the formation and production times is found to best fit the HERMES and EMC data (see Equation 3 and Fig.~1 of Ref.~\cite{gallmeister}). Note that the calculations  presented in \cite{gallmeister} are for $K^0$ and $\bar{K^0}$. For a given nucleus, we show the average of the above calculations to compare with the $K^0_s$ data. 

In some theoretical models~\cite{Accardi:2009review}, the decrease in \rha with $z$ can 
be explained by the struck parton losing energy before picking 
up a partner antiquark, on the way to forming a hadron.  One 
mechanism of energy loss is gluon radiation in the nuclear 
medium, although other mechanisms such as absorption~\cite{Accardi:2005} are also possible. This 
energy loss reduces the hadron energy $E_h$ and hence reduces 
$z$ for hadronization from a nucleus.  Because the fragmentation 
functions have a steep dependence on $z$, the ratio \rha 
generates a  steady decrease with $z$.
Other models \cite{Accardi:2009review} assume a decreased 
formation length for the struck quark to pick up a partner antiquark, 
followed by rescattering and absorption of the resulting hadron.  


The results for \ks at CLAS kinematics show  \rha for $z>0.6$ with the  Pb target to be below the HERMES results for charged kaons from xenon. One must be careful when 
comparing these two data sets because of the different kinematic ranges; 
however, the CLAS kinematics provide an additional point of constraint 
for theoretical models of hadronization.  Note that neutral kaons have 
a smaller reaction cross section than pions, and hence rescattering 
effects are minimized for $K^0$ hadronization. In particular, the $K^0$ data should be less affected by hadronic final-state interactions than pions, and more sensitive to QCD effects (such as gluon radiation) as the quark propagates through cold nuclear matter.

Fig.~\ref{fig:k0mult_pt2} shows the multiplicity ratios as a function 
of the transverse momentum squared.  At high $p_T^2$, there are very 
few $K^0_s$ events, leading to large statistical uncertainties.  
However, at small $p_T^2$ the statistics are reasonable and there is a 
clear target-dependence to the nuclear attenuation for $p_T^2 < 0.1$~ GeV$^2$. The $K^0_s$ multiplicity ratios then increase  as $p_T^2$  increases and exceed unity at around $p_T^2$=0.45 GeV$^2$. This crossover is at lower $p_T^2$ than for the HERMES results. Furthermore, the CLAS data have a steeper slope with $p_T^2$  than those measured by HERMES. Both CLAS and HERMES data show that \rha levels off for lighter targets at low $p_T^2$, yet continues to decrease for heavier targets at low $p_T^2$.

The increase in \rha with $p_T^2$ was seen previously by 
EMC~\cite{EMC_had} and HERMES~\cite{hermes2003,hermes2007}. This is 
known as the Cronin effect~\cite{cronin}, in which rescattering of either the parton or the hadron, as well as hadronic final state interactions, pushes events up to larger $p_T^2$. These effects occur preferentially in nuclear targets. The HERMES data for pions \cite{hermes2007, hermes_ptbroad2010} shows that the Cronin effect 
diminishes for high-$z$ events, which is consistent with predictions from 
parton rescattering \cite{kopeliovich}. The question  whether this also happens for \ks  hadronization must 
await better statistics, which will be available after the CLAS detector 
is upgraded in a few years for 12 GeV beam energies. 

In summary, we have extracted for the first time the multiplicity ratios 
for \ks using semi-inclusive deep inelastic scattering with detection 
of the \ks using  the CLAS detector with a 5.014 GeV electron beam incident on 
both deuterium and nuclear targets.  Many systematic effects cancel in 
the ratio $R^h_A$, in part because both deuterium and nuclear targets were exposed to beam simultaneously and were separated by only a few centimeters at the 
center of CLAS.   The $z$ and $p_T^2$ dependencies 
of \rha follow the trends seen in previous data by HERMES 
Collaboration  at higher $\nu$ and $Q^2$, 
suggesting that the general mechanism of hadron formation  are similar at CLAS kinematics.

In order to fully understand the fundamental principles of the hadron formation as well as the flavor dependence of the observables, a wide spectrum of both baryons and mesons are required. Some of these channels, including more precise pion data, are currently under investigation by the CLAS collaboration.  Ultimately, the interpretation of these data will require further theoretical development.

We thank the staff of the Accelerator and Physics Divisions at  Jefferson Lab for their support. This work was supported in part by the Chilean Comisi\'on Nacional de Investigaci\'on Cient\'ifica y Tecnol\'ogica (CONICYT), the Italian Istituto Nazionale di Fisica Nucleare, the French Centre National de la Recherche Scientifique, the French Commissariat \`{a} l'Energie Atomique, the U.S. Department of Energy, the National Science Foundation, the UK Science and Technology Facilities Council (STFC), the Scottish Universities Physics Alliance (SUPA), and the National Research Foundation of Korea. The Jefferson Science Assosciates (JSA) and Southeastern Universities Research Association (SURA) which operates the Thomas Jefferson National Accelerator Facility for the United States Department of Energy under contract DE-AC05-84ER40150.


\bibliographystyle{model1-num-names}
\bibliography{k0paper}







\end{document}